\documentclass[twocolumn,showpacs,preprintnumbers,amsmath,amssymb]{revtex4}
\usepackage{graphicx}% Include figure files
\usepackage{dcolumn}% Align table columns on decimal point
\usepackage{bm}% bold math

\begin{document}

\preprint{}

\title{Rotational Doppler effect in left-handed materials}% Force line breaks with \\
\author{Hailu Luo}
%\altaffiliation[ ]{}%Lines break automatically or can be forced with \\
\author{Shuangchun Wen}\email{scwen@hnu.cn}
\author{Weixing Shu}
\author{Zhixiang Tang}
\author{Yanhong Zou}
\author{Dianyuan Fan}
%%\email{Second.Author@institution.edu}
\affiliation{Key Laboratory for Micro/Nano Opto-Electronic Devices
of Ministry of Education, School of Computer and Communication,
Hunan University, Changsha 410082, China}
\date{\today}% It is always \today, today,

\begin{abstract}
We explain the rotational Doppler effect associated with light beams
carrying with orbital angular momentum in left-handed materials
(LHMs). We demonstrate that the rotational Doppler effect in LHMs is
unreversed, which is significantly different from the linear Doppler
effect. The physics underlying this intriguing effect is the
combined contributions of negative phase velocity and inverse screw
of wave-front. In the normal dispersion region, the rotational
Doppler effect induces a upstream energy flow but a downstream
momentum flow. In the anomalous dispersion region, however, the
rotational Doppler effect produces a downstream energy flow but a
upstream momentum flow. We theoretically predict that the rotational
Doppler effect can induce a transfer of angular momentum of the LHM
to orbital angular momentum of the beam.
\end{abstract}

%\keywords{}
%Use showkeys class option if keyword
\maketitle

\section{Introduction}\label{Introduction}
After the first experimental observation of negative
refraction~\cite{Smith2000,Shelby2001}, some intriguing or even
counter-intuitive phenomena in left-handed materials (LHMs), such as
amplification of evanescent waves~\cite{Pendry2000,Fang2005},
unusual photon tunneling~\cite{Zhang2002,Kim2004}, negative
Goos-H\"{a}nchen shift~\cite{Kong2002,Berman2002}, and reversed
linear Doppler effect~\cite{Seddon2003,Stancil2006} have attracted
much attention. Linear Doppler effect is a well-known phenomenon by
which the frequency of a wave is shifted according to the relative
velocity of the source and the observer~\cite{Jackson1999}. The
conventional understanding of linear Doppler effect is that
increased frequencies are measured when a source and an observer
approach each other. Applications of the effect are widely
established and include radar, laser cooling, flow measurement, and
the search for astronomical objects. The inverse linear Doppler
effect refers to frequency shifts that are in the opposite sense to
those described above. For example, increased frequencies would be
measured on reflection of waves from a receding
boundary~\cite{Seddon2003,Stancil2006}.

The interaction of a moving medium with a Lagurre-Gaussian (LG)
field raises the fundamental question of the Doppler effect. As a
medium moves across the helical wave fronts of the LG field, it
experiences, in addition to the usual linear Doppler shift related
to the velocity in the light propagation direction, a most
intriguing frequency shift, the so-called rotational Doppler effect
associated with the azimuthal velocity. The rotational Doppler
effect results in frequency shift when the light beam is rotated
around its propagation axis. The rotational Doppler effect has been
extensively studied in regular right-handed materials
(RHMs)~\cite{Garetz1979,Allen1994,Nienhuis1996,Courtial1998a,Courtial1998b}.
Veselago has suggested that LHMs would reverse nearly all known
optical phenomena~\cite{Veselago1968}. Now an interesting question
arises: what happens to the rotational Doppler effect in LHMs? When
the LG beams incident into LHMs, the helical wave fronts should
reverse their screwing fashion~\cite{Luo2008}. Thus we can predict
that the rotational Doppler effect will result in many intriguing
phenomena. The investigation of the rotational Doppler effect will
provide insights into the fundamental properties of LHMs and will
allow us to better understand the interaction of light with LHMs.

In this work, we try to reveal the rotational Doppler effect
associated with light beams carrying orbital angular momentum in
LHMs. First, starting from the Maxwell's equations in moving frame,
we obtain the analytical description for LG beams propagating in
rotating LHMs. Our formalism permits us to introduce the effective
index to describe the wave propagation. Next, we attempt to recover
how the wave-front evolves, and how the screwing wave-fronts result
in upshifted and downshifted frequencies. In order to explore the
rotational Doppler effect, we examine a simplest intensity pattern
formed by LG beams with equal magnitudes and waist parameters. Then,
we want to investigate how the rotational Doppler effect gives rise
to the anomalous upstream transverse Poynting vector and angular
momentum flow. Finally, we attempt to explore how the rotational
Doppler effect influences the rotation of intensity pattern inside
the LHM.

\section{Effective refractive index}\label{II}
To investigate the effective index of rotating LHM, we firstly use
the Maxwell's equations to determine the field distribution inside
the LHM. We consider a monochromatic electromagnetic field, ${\bf
E}({\bf r},t) = \text{Re} [{\bf E}({\bf r})\exp(-i\omega t)]$ and
${\bf B}({\bf r},t) = \text{Re} [{\bf B}({\bf r})\exp(-i\omega t)]$,
of angular frequency $\omega$ propagating from a conventional RHM to
the LHM. The field can be described by Maxwell's equations
\begin{eqnarray}
\nabla\times {\bf E} &=& - \frac{\partial {\bf B}}{\partial t},
~~~\nabla \cdot {\bf B} =0,\nonumber\\
\nabla\times {\bf H} &=& \frac{\partial {\bf D}}{\partial
t},~~~~~\nabla \cdot {\bf D} =0. \label{maxwell}
\end{eqnarray}
From the Maxwell's equations, we can easily find that the wave
propagation is permitted in materials with $\varepsilon<0$ and
$\mu<0$. Veselago termed these LHMs, because the vectors ${\bf E}$,
${\bf H}$ and ${\bf k}$ form a left-handed triplet instead of a
right-handed triplet, as is the case in regular
RHMs~\cite{Veselago1968}.

In order to investigate the effective index, we introduce a rest
frame and a moving frame~\cite{Player1976}. In the rest frame the
LHM rotates with an angular velocity ${\bf v}_\varphi={\bf
\Omega}\times{\bf r}$ and in the moving frame the LHM is at rest. We
restrict our analysis to a small velocity with $v\ll c$. For the LHM
at rest, we have the following constitutive relations:
\begin{equation}
{\bf D}'=\varepsilon_0 \varepsilon (\omega') {\bf E}'~~~~~~{\bf
B}'=\mu_0 \mu (\omega') {\bf H}'
\end{equation}
where we have used primes to denote the fields and their frequencies
in the moving frame. The fields in the moving frame can be expressed
in the rest frame by a Lorentz transformation, which gives to first
order in $v/c$:
\begin{eqnarray}
{\bf D}+\frac{{\bf v}\times{\bf H}}{c^2}&=&\varepsilon_0 \varepsilon
(\omega)
({\bf E}+{\bf v}\times{\bf B}),\\
{\bf B}-\frac{{\bf v}\times{\bf E}}{c^2}&=&\mu_0 \mu (\omega) ({\bf
H}-{\bf v}\times{\bf D}).\label{Efield}
\end{eqnarray}
Here electric permittivity $\varepsilon$ and magnetic permeability
$\mu$ are given as a function of the frequency in the moving frame.
We also assume that $\varepsilon$ and $\mu$ depend only on the
frequency and is independent of the state of motion of the LHM. On
combining these two equations, we can obtain the following wave
equation:
\begin{equation}
\nabla^2 D=-n(\omega'_{\pm})^2\omega^2
D+2[n(\omega'_{\pm})^2-1]\Omega\omega(\sigma\pm
\partial_\varphi)D,\label{WV}
\end{equation}
where we have introduced the relation
$n(\omega')=-\sqrt{\varepsilon(\omega')\mu(\omega')}$, and
$\sigma=\pm1$ corresponds to left- and right-handed circularly
polarized light, respectively. To determine the effective indices,
we solve the wave equation for helical wave-front LG beams.

We introduce the Lorentz-gauge vector potential to describe the
propagation characteristics of LG filed. The vector potential of the
beam propagating in the $+z$ direction can be written in the form
\begin{equation}
{\bf A}=A_0(\alpha{\bf e}_x+\beta{\bf e}_y)u_{p,l}({\bf r})\exp(i n
k_0 z-i\omega t).\label{ca}
\end{equation}
Here $A_0$ is a complex amplitude, $n$ is the effective index,
$k_0=\omega/c$, $c$ is the speed of light in vacuum, ${\bf e}_x$ and
${\bf e}_y$ are unit vectors, respectively. The coefficients
$\alpha$ and $\beta$ satisfy the relation
$\sigma=i(\alpha\beta^\ast-\alpha^\ast\beta)$. It is well known that
each photon in a light beam carries a spin angular momentum.

\begin{figure}
\includegraphics[width=6cm]{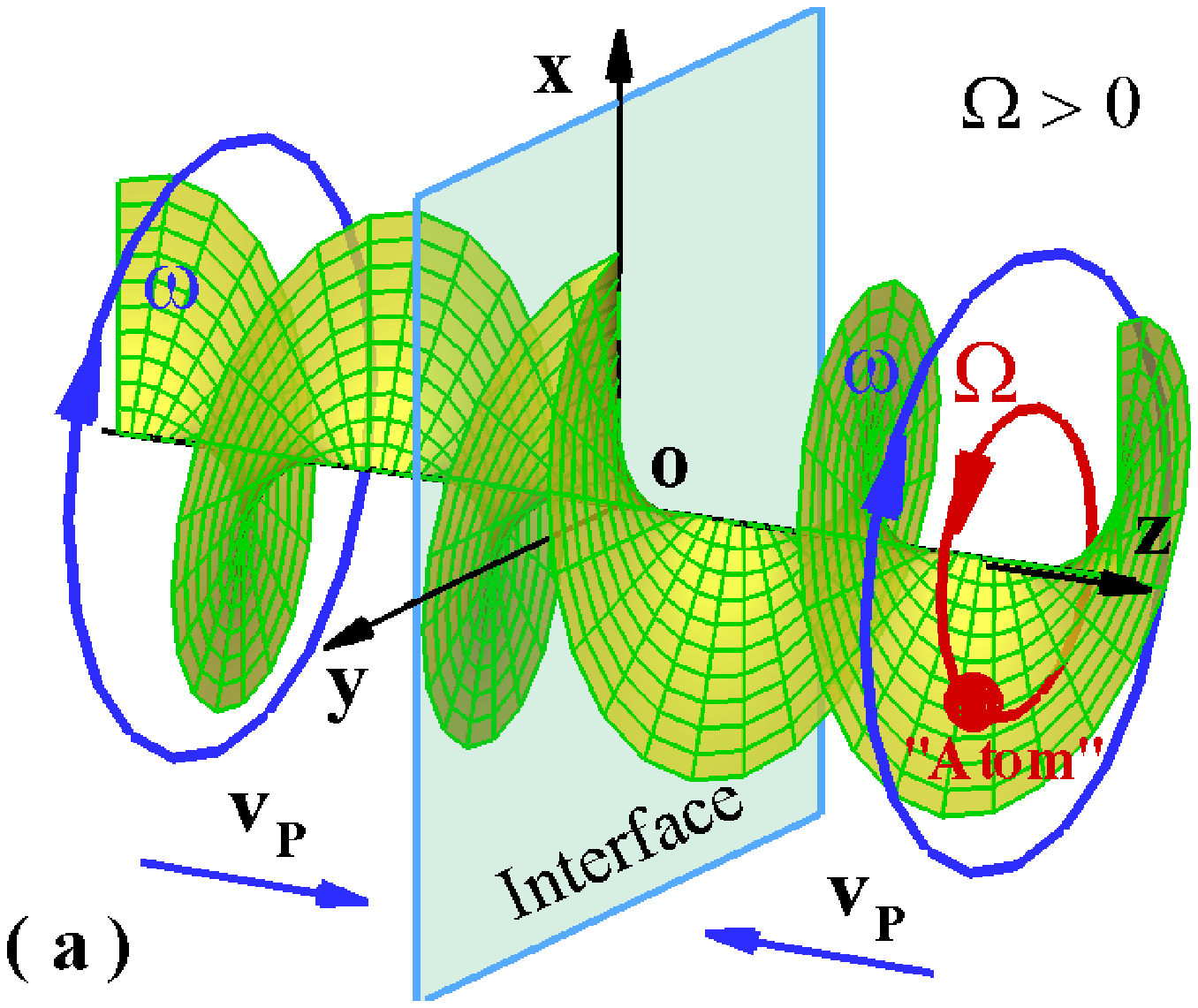}
\includegraphics[width=6cm]{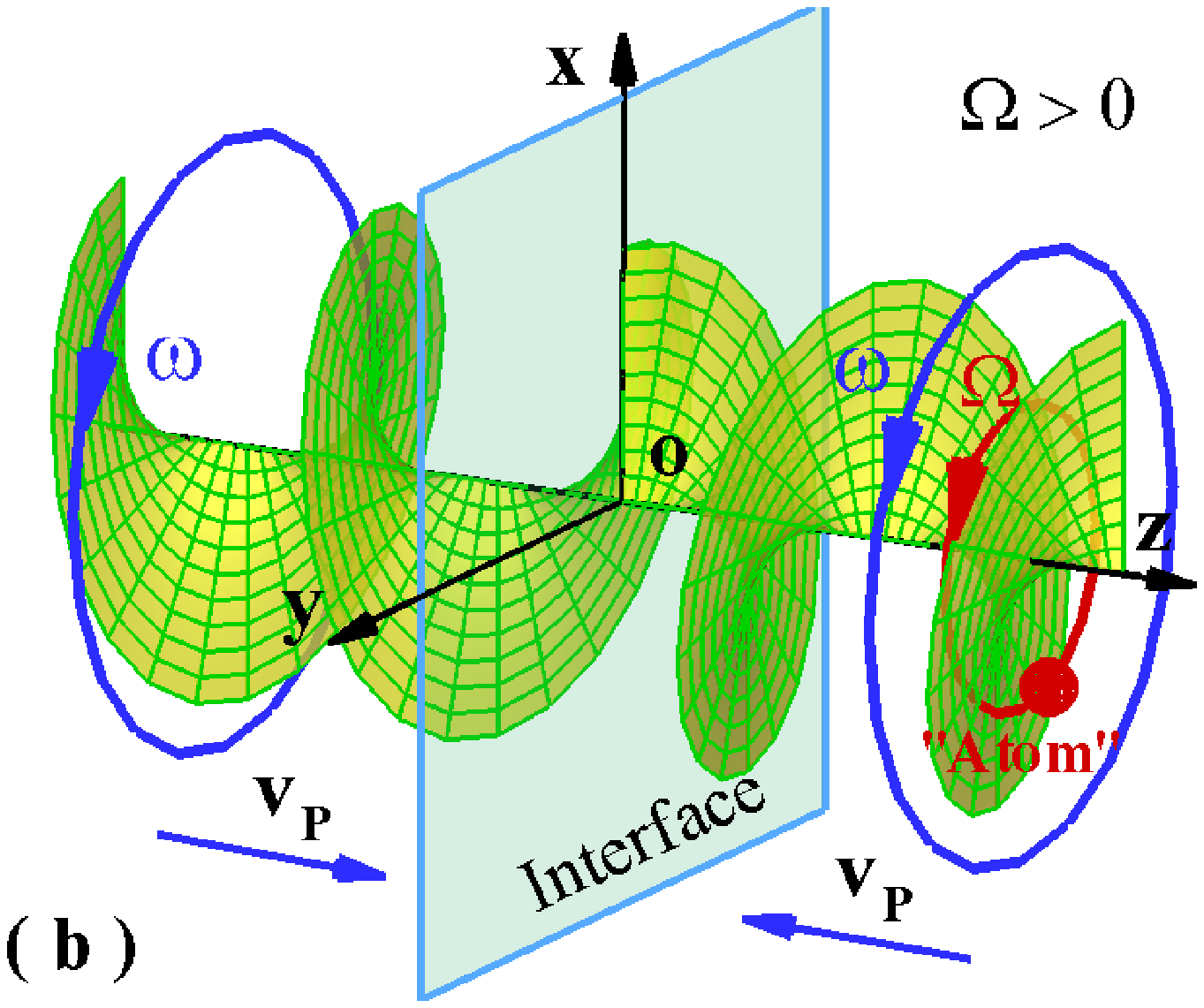}
% Here is how to import EPS art
\caption{\label{Fig1} (Color online) The helical wave front for
Laguerre-Gaussian beam result from an azimuthal phase structure of
$\exp[\pm il\varphi]$. The LHM is rotating with angular velocity
$\Omega$. The phase velocity ${\bf v}_p$ reverses its direction and
the wave-fronts reverse their screwing rotation in the LHM. When a
metamaterial ``atom'' moves across the helical wave fronts, it
experiences the so-called rotational Doppler effect. (a) The helical
wave front with $\exp[+i\varphi]$ and the effective frequency
$\omega'=\omega+\Omega$. (b) The helical wave front with $\exp[-
i\varphi]$ and the effective frequency $\omega'=\omega-\Omega$.}
\end{figure}

To be uniform throughout the following analysis, we introduce
different coordinate transformations $z_i^\ast (i=1,2)$ in the RHM
and the LHM, respectively. The field can be written as
\begin{eqnarray}
u_{pl}=&&\frac{C_{pl}}{w(z_i^\ast)}
\left[\frac{\sqrt{2}r}{w^2(z_i^\ast)}\right]^{l}
L_p^{l}\left[\frac{\sqrt{2}r}{w^2(z_i^\ast)}\right]
\exp\bigg[\frac{-r^2}{w^2(z_i^\ast)}\bigg]\nonumber\\
&&\times \exp\bigg[\frac{-i n k_0 r^2
z_i^\ast}{R(z_i^\ast)}\bigg]\exp[\pm il\varphi]\nonumber\\ &&\times
\exp[-i (2p+l+1)\arctan (z_i^\ast/z_{R})],\label{F1}
\end{eqnarray}
\begin{eqnarray}
w(z_{1}^\ast)=w_0\sqrt{1+(z_{i}^\ast/z_{R})^2},~~R(z_{i}^\ast)=z_i^\ast+\frac{z_{R}^2}{z_i^\ast}.
\end{eqnarray}
Here $C_{pl}$ is the normalization constant, $L_p^l[2
r^2/w^2(z_1^\ast)]$ is a generalized Laguerre polynomial, $z_{R}= n
k_0 w_0^2 /2$ is the Rayleigh length, and $w(z_{i}^\ast)$ is the
beam size and $R(z_{i}^\ast)$ the radius of curvature of the wave
front. The last term in Eq.~(\ref{F1}) denotes the Gouy phase which
is given by $\Phi=-(2p+l+1)\arctan(z_i^\ast/z_{R})$.

Now let us study the screw of the wave front. Here the sense of the
positive angles is chosen as anticlockwise, while negative angles
are considered in the clockwise direction. In the rotating LHM, the
constant wavefront satisfies
\begin{eqnarray}
n_{+} k_0 z_i^\ast+\frac{-i n_+ k_0 r^2
}{R_+(z_i^\ast)}+l\varphi+\Phi_+(z_i^\ast)=\text{const},\label{PP}\\
n_{-} k_0 z_i^\ast+\frac{-i n_- k_0 r^2
}{R_-(z_i^\ast)}-l\varphi+\Phi_-(z_i^\ast)=\text{const}.\label{PN}
\end{eqnarray}
The schematic view of the wave front is a three-dimensional screw
surface of ($r\cos\varphi$, $r \sin\varphi$, $z$). The plotting
range of $r$ is from $0$ to $5w_0$ with the interval of $\Delta
r=0.5 w_0$ and that of $|n_\pm| k_0 z$ is from $0$ to $4\pi$ with
the interval of $|n_\pm| k_0 \Delta z=0.1\pi$. The wavefront
structure reverse their screw types with a pitch of
$\lambda_0/|n_+|$ and $\lambda_0/|n_-|$, respectively (see
Fig.~\ref{Fig1}).

The helical wave fronts with clockwise or anticlockwise screwing
fashion will result in different effective indices. Substituting the
LG beam in the wave equation, we have
\begin{equation}
k_{\pm}^2(\omega)=n^2(\omega'_{\pm})\frac{\omega^2}{c^2}-2[n^2(\omega'_{\pm})-1]
\frac{\Omega\omega}{c^2}(\sigma\pm l).\label{WV}
\end{equation}
Thus from Eq.~(\ref{WV}), and using the relation
$k(\omega)=n(\omega')\omega/c$, we can obtain an equation of the
refractive indices associated with the waves of each screwing
fashion:
\begin{equation}
n_{\pm}(\omega)=n(\omega'_{\pm})-\left[n(\omega'_{\pm})-
\frac{1}{n(\omega'_{\pm})}\right]\frac{\Omega}{\omega}(\sigma\pm
l).\label{Efield}
\end{equation}
Here we have approximated the refractive indices to the first order
in $\Omega/\omega$. To obtain the indices, we need to know the
effective frequency. Strictly speaking, we cannot obtain it from the
wave equation in the moving frame. Now a question arises: how to
determine the value of effective frequency in the LHM?

\section{Rotational Doppler effect}\label{III}
In the case of electromagnetic radiation this usually means that the
subunits must be much smaller than the wavelength of radiation. Then
the unit cells of metamaterials can be modeled as the atoms (or
molecules) in ordinary materials. In particular, electric atoms with
an electric-dipole moment leading to a negative electric
permittivity, and magnetic atoms with a magnetic-dipole moment
leading to a negative magnetic permeability. Hence the rotational
Doppler effect in LHMs is the result of the optical properties of
the individual metamaterial ``atoms'' (see Fig.~\ref{Fig1}).

To quantify the changes that occur in the Doppler effect for
metamaterial ``atoms'' interacting with light beams with orbital
angular momentum, the quantum mechanical derivation~\cite{Allen1994}
is developed for describing the rotational Doppler effect in LHMs.
The Doppler shift experienced by a LG field in a moving frame is
given by
\begin{eqnarray}
\delta_{LG}&=&\bigg[-n k_0+ \frac{n k_0
r^2}{2(z^2_i+z_R^2)}\left(\frac{2z^2_i}{z^2_i+z_R^2}-1\right)\nonumber\\
&&-\frac{(2p+l+1)z_R}{z^2_i+z_R^2}\bigg]v_z -\frac{n k_0 r}{R}
v_r-\frac{\sigma\pm l}{r}v_\varphi,\label{Dop}
\end{eqnarray}
where $v_z$, $v_r$, and $v_\varphi$ are the axial, radial, and
azimuthal velocity components of the metamaterial ``atom'',
respectively. The axial Doppler shift is obtained from $n k_0 v_z$.
There are two additional terms which account for the radial phase
and the Gouy-phase. Note that the two additional terms should be
reversed, since the negative index and the inverse Gouy-phase
shift~\cite{Luo2007}. Thus we can conclude that the linear Doppler
effect associated with light beams is slightly different form the
counterpart of plane wave~\cite{Seddon2003,Stancil2006}. Further
investigation shows that the radial Doppler shift, given by the term
proportional to $v_r$, should also be reversed in LHMs.

A most intriguing frequency shift, the so-called rotational Doppler
effect arises from the azimuthal velocity $v_\varphi=\Omega r$. The
rotational Doppler effect results in frequency shift when the light
beam is rotated around its propagation axis. The corresponding
rotational Doppler shift in LHMs is given by
$\delta_{LG}^\varphi=(\sigma\pm l)\Omega$, which is proportional to
the total angular momentum. It can be seen that beams propagate
differently in a rotating LHM, depending on whether the wave fronts
turn in the same rotation sense as the LHM or in the opposite sense.
The frequency of an incident LG beam whose wave-front is screwing as
the LHM is downshifted to $\omega'=\omega-(\sigma+l)\Omega$, whereas
a LG beam of the opposite sense is upshifted to
$\omega'=\omega-(\sigma-l)\Omega$. Unexpectedly, the rotational
frequency shift in LHMs is unreversed.

When a LG beam incident into a LHM, the helical wave fronts reverse
their screwing fashion~\cite{Luo2008}. Why is the rotational
frequency shift unreversed? By closely examining the evolution of
wave-front, we find the phase velocity also inverses its direction
(see Fig.~\ref{Fig1}). Hence the dynamic evolution between the
metamaterial ``atom'' and wave-front remains unchanged. We conclude
that the physics underlying the unreversed effect is collective
contributions of negative phase velocity and inverse screw of
wave-front.

Though the rotational Doppler effect is unreversed, much more
counter-intuitive phenomena will be caused in the LHMs. In order to
explore the rotational Doppler effect, we will examine a simplest
intensity pattern which is formed by $\text{LG}_{0,+1}$ and
$\text{LG}_{0,-1}$ beams with equal magnitudes and waist parameters.
In principle, any arbitrary light beam can be described by a
superposition of LG modes~\cite{Allen2003}. In the case of equal
frequencies, such a combination is well known to be equivalent to an
ordinary Hermite-Gaussian (HG) beam with a double-spot intensity
distribution. We will explore the energy flow and angular momentum
flow caused by rotational Doppler effect.

First let us examine the propagation characteristics of energy flow,
which is usually discussed by use of the Poynting vector. There has
been considerable interest in orbital angular momentum of LG beams
relating to azimuthal component of Poynting
vector~\cite{Allen1992,Padgett1995,Allen2000}. The time average
Poynting vector, ${\bf S}$, can be written as
\begin{equation}
{\bf S}=\frac{1}{2}\text{Re}[({{\bf E}_+}+{{\bf E}_-})\times({{\bf
H}_+}+{{\bf H}_-})^\ast].\label{PV}
\end{equation}
We denote the radial, azimuthal, and axial components  of the vector
${\bf S}$ in the cylindrical coordinates by the notation $S_r$,
$S_\varphi$, and $S_z$, respectively. The component $S_r$, relates
to the spread of the beam as it propagates. The azimuthal component
$S_\varphi$ describes the energy flow that circulates around the
propagating axis. The axial component $S_z$ describes the energy
flow that propagates along the $+z$ axis.

\begin{figure}
\includegraphics[width=7cm]{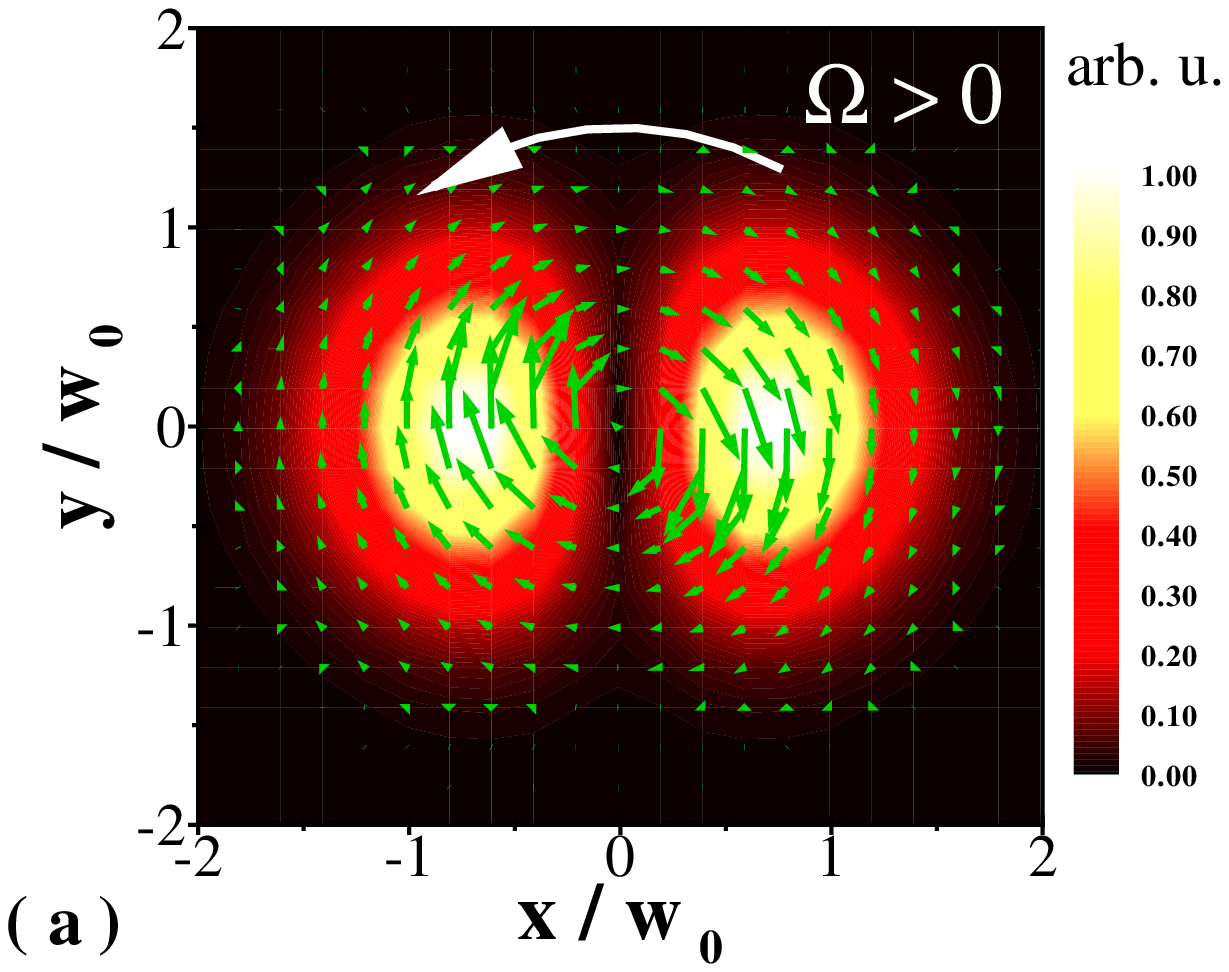}
\includegraphics[width=7cm]{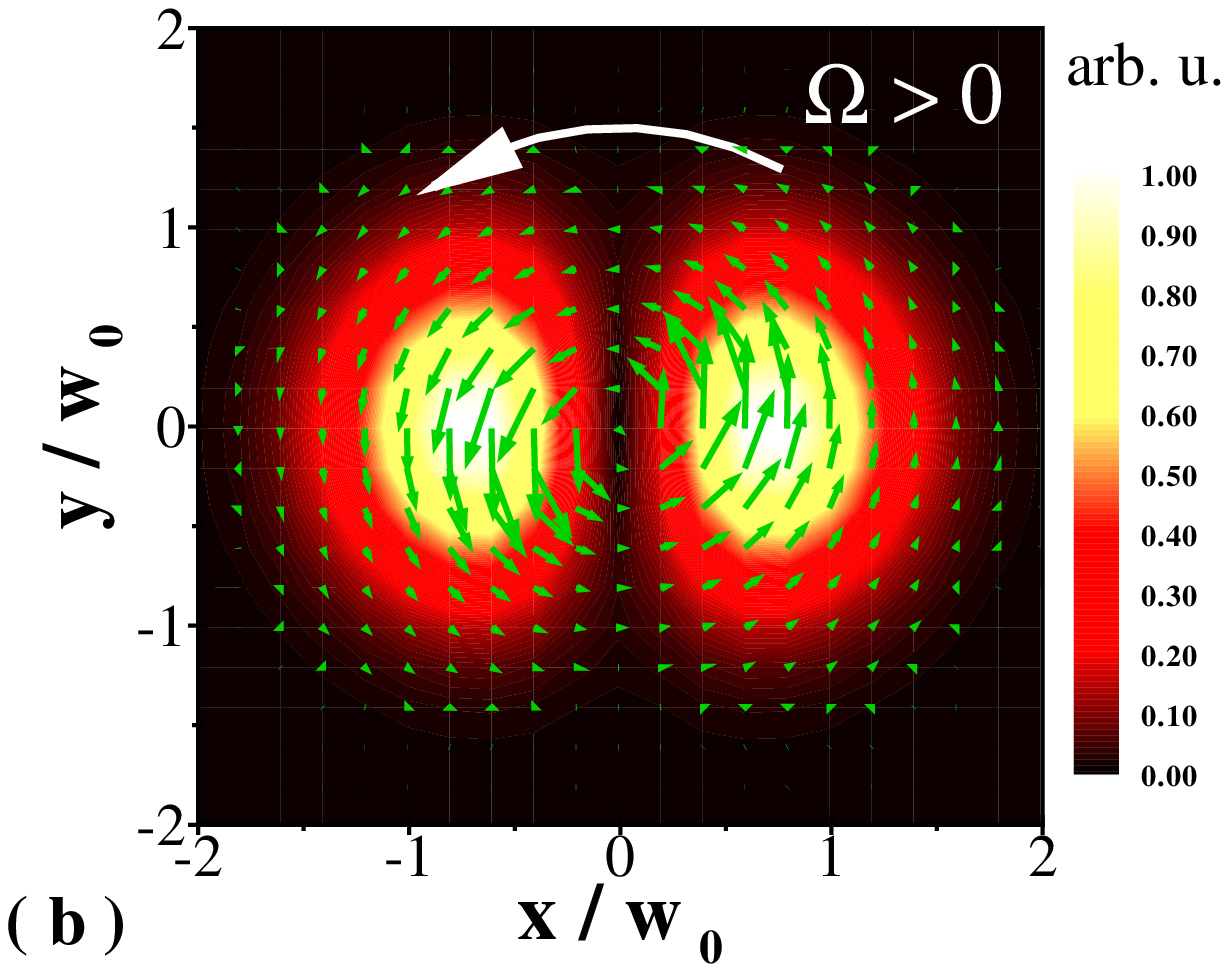}
% Here is how to import EPS art
\caption{\label{Fig2} (Color online) In the normal dispersion region
, the rotational Doppler effect induces (a) the downstream energy
flow and (b) the anomalous upstream transverse momentum flow. The
parameters are $\omega=1.366\omega_0$ and $z_2^\ast=0$. The green
arrows denotes the transfers Poynting vector and momentum flow.
Please note that the beam divergence is not shown due to
normalization of transverse dimension.}
\end{figure}

Next we attempt to describe the evolution of momentum flow. The
momentum flow for the electromagnetic wave can be written in the
form~\cite{Jackson1999}
\begin{equation}
{\bf T}=\frac{1}{2}\text{Re} [({\bf D}\cdot{\bf E}^\ast+{\bf
B}\cdot{\bf H}^\ast)I -({\bf D} {\bf E}^\ast+{\bf B} {\bf H}^\ast)].
\end{equation}
Here ${\bf T}$ is the momentum flow also referred as the Maxwell
stress tensor, and $I$ is $3\times3$ identity matrix. For a HG beam,
${{\bf E}}={{\bf E}_+}+{{\bf E}_-}$ and ${{\bf H}}={{\bf H}_+}+{{\bf
H}_-}$. Note that the momentum flow is significantly different from
that obtained by Minkowski or Abraham momentum~\cite{Jackson1999}.
The cross product of this momentum density with the radius vector
yields an orbital angular momentum flow. The orbital angular
momentum flow in the $z$ direction depends upon the component of
${\bf T}_\varphi$, such that ${\bf J}_z= {\bf r} \times {\bf
T}_\varphi$.

In order to accurately describe the Poynting vector and momentum
flow, it is necessary to include material dispersion and
losses~\cite{Kemp2007}. Thus, a certain dispersion relation, such as
the Lorentz medium model, should be introduced. The constitutive
parameters are
\begin{eqnarray}
\varepsilon(\omega)&=&\varepsilon_0\left(1-\frac{\omega_{ep}^2}{\omega^2
-\omega_{eo}^2+i\omega \gamma_e}\right),\\
\mu(\omega)&=&\mu_0\left(1-\frac{F\omega_{mp}^2}{\omega^2-\omega_{mo}^2+i\omega
\gamma_m}\right).
\end{eqnarray}
To avoid the trouble involving in a certain value of frequency, we
assume the material parameters are
$\omega_{eo}=\omega_{mo}=\omega_o$, $\omega_{ep}^2=F\omega_{mp}^2=2
\omega_o^2$ and $\gamma_e=\gamma_m=0.25\omega_0$. It should be noted
that, as we will see in the following, the dispersion and negative
indices will play a very important role in the rotation of Poynting
vector and momentum flow.

Figure~\ref{Fig2} shows the characteristics of Poynting vector and
momentum flow. The circumstance is really intriguing: the novel
azimuthal energy flow and momentum flow present in the HG field.
Counter-intuitively, the angular velocity $\Omega$ of the rotating
LHM and the circulation of Poynting vector are opposite, as shown in
Fig.~\ref{Fig2}(a). The presence of the azimuthal momentum flow
${\bf T}_\varphi$ has been identified as the orbital angular
momentum. We thus predict that the rotational Doppler effect can
induce a transfer of angular momentum of the LHM to obitual angular
momentum of the beam. The sign of the orbital angular momentum
coincides with handedness of rotation of the LHM. Positive orbital
angular momentum is associated with the counter-clockwise rotation
$\Omega>0$, as plotted in Fig.~\ref{Fig2}(b). On the contrary, the
sign of the orbital angular momentum in the HG field is negative:
when the LHM rotates clockwise $\Omega<0$, the angular momentum flow
seemingly corresponds to the clockwise circulation. The physical
origin of transfer of angular momentum is rooted in the geometric
phase difference between $\text{LG}_{0,+1}$ and $\text{LG}_{0,-1}$
beams.

Whether the rotational Doppler effect induce the downstream or
upstream angular momentum depends on the normal or anomalous
dispersion. In the normal dispersion region, $\partial
\text{Re}[n(\omega)]/\partial\omega>0$, the rotational Doppler
effect induces a upstream energy flow [see Fig.~\ref{Fig2}(a)] but a
downstream momentum flow [see Fig.~\ref{Fig2}(b)]. In the anomalous
dispersion region, $\partial \text{Re}[n(\omega)]/\partial\omega<0$,
however, the rotational Doppler effect induces a downstream energy
flow  and a upstream momentum flow. Interestingly, we find that the
transverse energy flow in the LHM is antiparallel to the transverse
momentum flow. In our opinion, the main reason for such an
inconsistence is the negative refractive index. Note that the novel
phenomenon is significantly different from the counterpart in
regular RHMs~\cite{Bekshaev2005}.

\begin{figure}
\includegraphics[width=7cm]{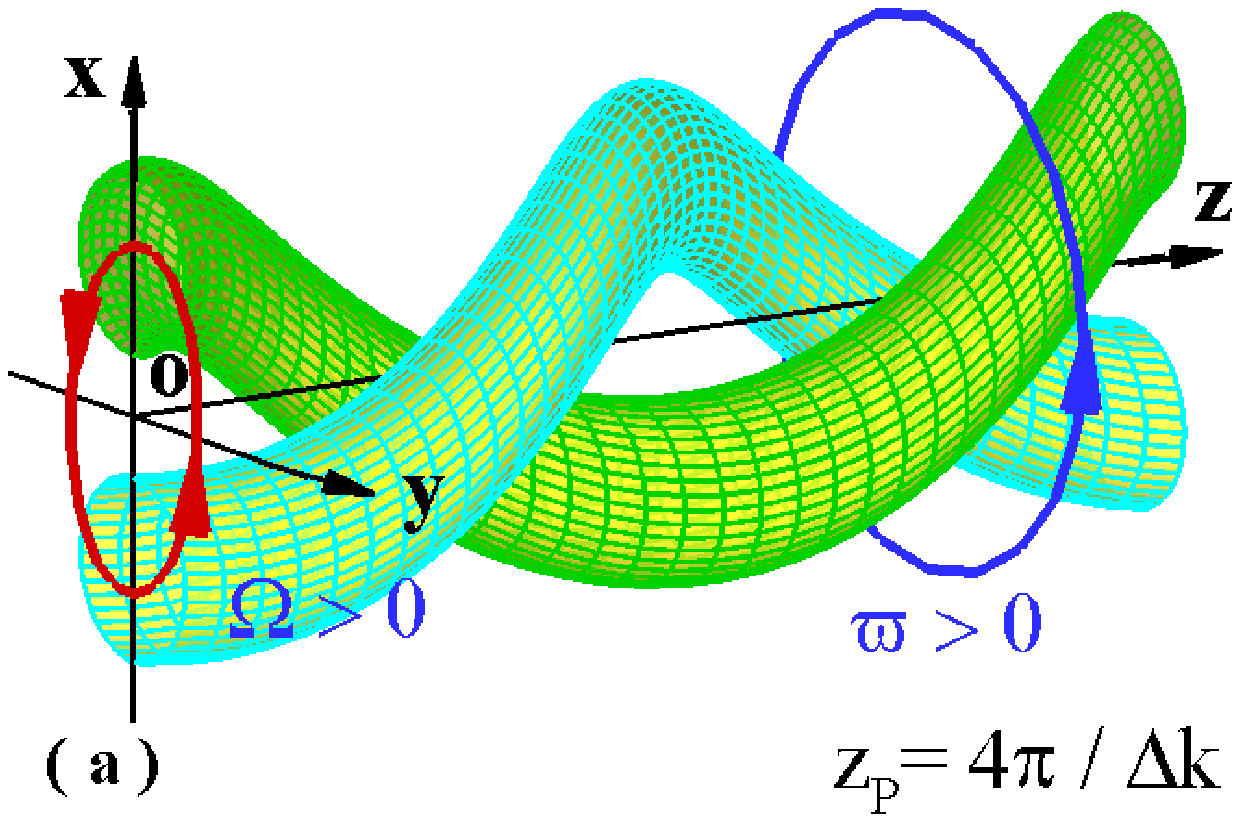}
\includegraphics[width=7cm]{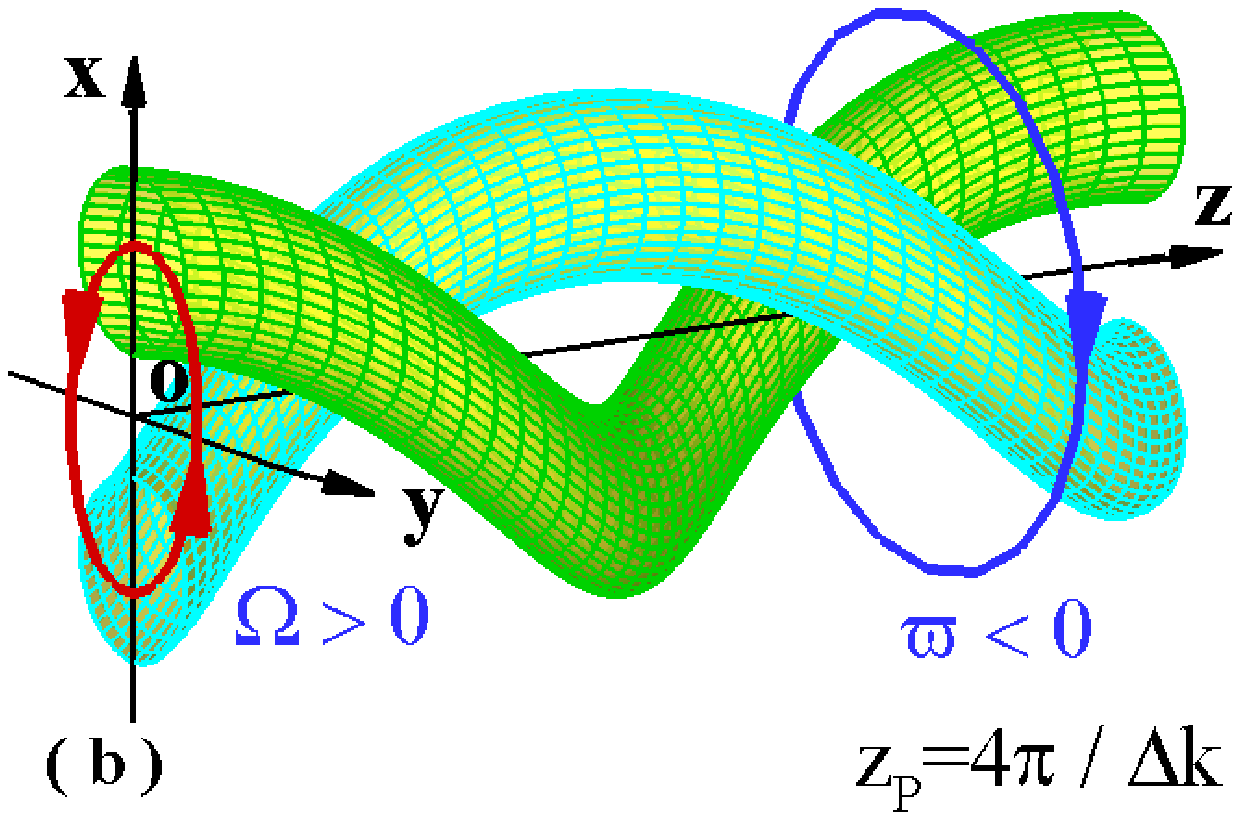}
% Here is how to import EPS art
\caption{\label{Fig3} (Color online) Instant spatial structure of HG
beam screws with pitch $z_\text{p}=4\pi/(k_+-k_-)$ in the LHM. The
rotational Doppler effect induces rotation of intensity pattern: (a)
the downstream rotation ($\Omega>0$, $\varpi>0$) in normal
dispersion region, (b) the upstream rotation ($\Omega>0$,
$\varpi<0$) in anomalous dispersion region. Please note that the
beam divergence is not shown due to normalization of transverse
dimension. }
\end{figure}

To understand a better physical picture, let us reveal the instant
field distribution in the LHM. The rotation of the intensity pattern
in LHM arises from the difference in the refractive indices for
constituent LG beams. The transverse beam pattern is determined by
the geometric phase difference:
\begin{equation}
\Delta k z_i^\ast+\frac{k_+ r^2 }{R_+(z_i^\ast)}-\frac{k_- r^2
}{R_-(z_i^\ast)}+2l\varphi-\Delta \Phi=\text{const},
\end{equation}
where $\Delta k=k_+ - k_-$, $k_\pm=n_\pm(\omega')\omega/c$, and
$\Delta \Phi=\Phi_+ - \Phi_-$. The schematic view of the spatial
structure of instant field is plotted in Fig.~\ref{Fig3}. Here we
have neglected the phase difference caused by radial and Gouy
phases. The spatial structure exhibits an screwing type with a pitch
of $z_\text{p}=4\pi/(k_+-k_-)$ along the $+z$ axis.

In this case, the constituent LG beams propagate differently in the
LHM which leads to a change in their relative phase. The pattern
rotates with an angular velocity $\varpi=c(k_+-k_-)/2$. The angle
per unit length by which the intensity pattern is rotated is called
the specific rotary power
$\Delta\theta_{\text{I}}=2\pi(n_+-n_-)/\lambda$. Hence the
rotational Doppler effect induces a downstream or a upstream
rotation of the intensity pattern, depending on whether the
frequency locates in the region of normal or anomalous dispersion.
In the normal dispersion region, $k_+ > k_-$, the angular velocity
$\varpi>0$, and the rotational Doppler effect induces a downstream
rotation of intensity pattern [see Fig.~\ref{Fig3}(a)].  In the
anomalous dispersion region, $k_+ < k_+$, the angular velocity
$\varpi<0$, and the rotational Doppler effect induces a downstream
rotation of intensity pattern [see Fig.~\ref{Fig3}(b)]. A further
point should be mentioned that the large rotation can be obtained if
$k_+$ and $k_-$ have opposite signs (i.e., $k_+>0$, $k_-<0$ or
$k_+<0$, $k_->0$ ). Our result suggests that the interesting images
rotation~\cite{Padgett2006,Gotte2007,Leach2008} might more readily
be accessed in LHMs.

The possibility of having an upstream rotation of intensity pattern
in the presence of anomalous dispersion, i.e., in the presence of
negative group velocity. As negative group velocities in dispersive
media are forbidden by Kramers-Kronig relations in all frequency
regions~\cite{Jackson1999}. Thus negative group velocities are
possible only in the presence of absorption. In a recent
experiment~\cite{Dolling2006}, the negative group velocity have been
demonstrated by observing that the pulse advances in time with
respect to the same wave packet propagating in vacuum. The low-loss
LHMs are therefore good candidates for the experimental observation
of significant upstream rotation of intensity pattern. Conversely,
the anomalous upstream rotation of intensity pattern could provide
an interesting way to reveal the nature of LHMs.

\section{Conclusions}
In conclusion, we have investigated the rotational Doppler effect
associated with light beams carrying orbital angular momentum in
LHMs. Unlike the reversed linear Doppler effect, the rotational
Doppler effect in LHMs is unreversed. We have revealed that the
physics underlying this intriguing effect is combined contributions
of negative phase velocity and inverse screw of wave-front. Though
the rotational Doppler effect is independent of the refractive
index, much more counter-intuitive phenomena will be caused in the
LHMs. In the normal dispersion region, the rotational Doppler effect
induces a downstream energy flow but a upstream momentum flow. In
the anomalous dispersion region, however, the rotational Doppler
effect induces a upstream energy flow but a downstream momentum
flow. We theoretically predict that the rotational Doppler effect
can induce a transfer of angular momentum of the LHM to orbital
angular momentum of the beam. It is possible that the study of
Poynting vector and momentum flow in LHMs may make a useful
contribution to long established Abraham-Minkowski dilemma. We have
shown that the momentum flow in LHMs should be antiparallel to the
Poynting vector. Thus the direction of momentum flow is
significantly different from that obtained by Minkowski or Abraham
momentum. We consider that this difference might help us to
illustrate this long established problem.

\begin{acknowledgements}
This work was supported by projects of the National Natural Science
Foundation of China (Grants Nos. 10576012, 10674045, and 60538010).
\end{acknowledgements}

\end{document}